\documentclass[12pt]{iopart}

\usepackage{graphicx}  
\usepackage{color}
\begin{document}

\title[]{High-order harmonic generation driven by chirped laser pulses induced by linear and non linear phenomena}

\author{E Neyra$^1$, F Videla$^1$, J A P\'erez-Hern\'andez$^2$, M F Ciappina$^3$,L Roso$^2$ and G A Torchia$^1$}

\address{$^1$Centro de Investigaciones \'Opticas (CIOp) CONICET La Plata-CICBA, Camino Centenario y 506, M.B. Gonnet, CP 1897, Provincia de Buenos Aires, Argentina}
\address{$^2$Centro de L\'aseres Pulsados (CLPU), Parque Cient\'{\i}fico, E-37008 Villamayor, Salamanca, Spain } 
\address{$^3$Max Planck Institute of Quantum Optics, Hans-Kopfermann Str.~1, D-85748 Garching, Germany}

\ead{enriquen@ciop.unlp.edu.ar}

\begin{abstract}

We present a theoretical study of high-order harmonic generation (HHG) driven by ultrashort optical pulses with different kind of chirps. The goal of the present work is perform a detailed study to clarify the relevant parameters in the chirped pulses to achieve a noticeable cut-off extensions in HHG. These chirped pulses are generated using both linear and nonlinear dispersive media.The description of the origin of the physical mechanisms responsible of this extension is, however, not usually reported with enough detail in the literature. The study of the behaviour of the harmonic cut-off with these kind of pulses is carried out in the classical context, by the integration of the Newton-Lorentz equation complemented with the quantum approach, based on the integration of the time dependent Schr\"odinger equation in full dimensions (TDSE-3D).

\end{abstract}


\vspace{2pc}


\maketitle

%

\section{Introduction}

The interaction of ultra-short intense laser with atoms or molecules triggers several nonlinear phenomena, among these, we can point out the high-order-harmonic-generation (HHG) process~\cite{1,2} as one of the most prominent.  HHG is a well-known phenomenon commonly used to generate coherent radiation in the range of extreme ultraviolet (XUV) to Soft-X-Ray spectral range. A simple and intuitive way to describe the underlying physical mechanism behind HHG in atoms and molecules, has been well established in the so-called three-step model~\cite{3,4,5,maciej} that can be briefly summarized as follows. In the first step an electronic wave packet is sent to the continuum by tunnel ionization through the potential barrier of the atom, which is a consequence of the non-perturbative interaction between the atom and the laser electric field. Secondly, the emitted electronic wave packet propagates in the continuum to be finally driven back when the laser electric field changes its sign, and, finally, this electronic wave packet has certain probability to recombine with the ion core, taking place the transformation of the excess of kinetic energy in high-harmonic photons (this last step is also known as recombination).

There exists two fundamental ways for the control of the high-harmonic radiation emitted by atoms and molecules. We can control either the temporal evolution of the driving electric field-envelope and carrier frequency~\cite{amelleprx,vozzi1}, or to manipulate the spatial properties or the driven laser field in a broad sense, e.g.~by including medium engineering and geometric effects~\cite{6,7} or by using spatial inhomogeneous fields to drive not only the HHG phenomenon~\cite{optexp2012,joseprl,classtrong}, but also the ATI electrons~\cite{atispra,atislpl}. Related to the carrier frequency, it is well known that one of the most important tools in the study of the spectral characteristics of HHG is the control of the chirp in the driven laser pulse. This is so because the HHG strongly depends on this parameter~\cite{8,9}. It was already established that the control and shape optimization of the driven pulse are the main points to take into account. Another important aspect to be considered, is the resolution and efficiency of the harmonic yield~\cite{8,9,10,11,12}. In some of these works the method used to control the chirping was to achieve a suitable separation distance between the diffracting gratings of the compressor. This kind of chirped pulses is similar to the one obtained in a dispersive media and its magnitude is proportional to the group velocity dispersion (GVD)~\cite{13}. Theoretical studies of HHG employing chirped pulses, within the framework of the single atom model, show it is possible to extend considerably the HHG cutoff. An additional interesting feature appears: the harmonic spectra present a clear continuum shape being the latter an essential property for the production of isolated attosecond pulses~\cite{14,15,16,17}. There is one point to emphasize, however: the theoretically proposed chirped pulses differ from the experimental ones because the chirps are nonlinear in nature and it is not possible to achieve this kind of pulses with only linear dispersive media. It is worth to mention that the spectral properties of all the proposed pulses change according to the chirp parameters. This mean that new frequency components will appear, both in the Fourier transform of the pulse and in the HHG spectrum. 

In this paper we discuss under which general conditions a typical femtosecond chirped pulse, should extend the cut off the harmonic spectrum. Our model is based on single atom simulations using the time-dependent Schr\"odinger equation in full dimensions (TDSE-3D). Related to the latter point, up to now there is not a well-established explanation about what type of chirp and envelope are really able to produce an extension of the HHG spectra~\cite{14,15,16,17}. Note that by type of chirp we understand the functional dependence of the carrier frequency with respect to the time. After defining this dependence, it will be possible to examine the frequency content in the pulse analysing its influence in the HHG process. Our predictions are in agreement with the general relationship $\omega_{cut-off} \propto I\lambda^2$ between the intensity, $I$, and wavelength of the driven pulse, $\lambda$, and the cut off frequency, $\omega_{cut-off}$, of the harmonic spectrum.
In order to effectively calculate the HHG spectra driven by chirped pulses, obtained by linear or nonlinear processes in the medium, we use them as a input in the TDSE-3D. Following the quantum simulation, the results are compared with the classical model showing an excellent agreement. The classical model appears to be instrumental in order to understand the underlying physics behind the HHG cutoff extension. Then different strategies are investigated to modify the type of chirp of the driven pulse, mainly considering the group velocity dispersion (GVD) effects and the utilization a functional dependence of non linear character.

\section{Theoretical Methods}


According to the three step model~\cite{3,4,maciej} the maximum photon energy, $E_{cut-off}$, in the harmonic spectrum is given by the classical cut-off law, 
\begin{equation}
E_{cut-off}=I_p+3.17U_{p}  
\label{eq:eq1}
\end{equation}
(atomic units are used throughout this paper unless otherwise stated) where $I_p$ is the ionization potential of the corresponding target atom or ion (in this work we will focus on the helium atom, $I_p=0.9$ a.u., i.e. $24.7$ eV), $\omega_0$ is the central laser frequency and $U_{p}$ is the ponderomotive energy given by:  
\begin{equation}
U_p=\frac{E_0^2}{4\omega_0^2}  
 \label{eq:eq2}
\end{equation}
being $E_0$ the peak amplitude of the laser electric field.
For chirped pulses the laser pulse frequency is time-dependent, consequently, the instantaneous ponderomotive energy is now given by,
\begin{equation}
U_p(t)=\frac{E(t)^2}{4\omega(t)^2}   
\label{eq:eq3}
\end{equation}
In the following, and in order to avoid ambiguities, we will denote by $U_{po}\equiv U_p=\frac{E_0^2}{4\omega_0^2}$ the quantity reported by the equation (\ref{eq:eq2}) and $U_p(t)$ the one given by the equation (\ref{eq:eq3}).

Therefore, according to the equation (\ref{eq:eq3}), $ U_p(t)$  will take values lower, equal or higher than $U_{po}$ for certain time intervals $\Delta t$. In order to determine this relationship, we will define the following function,
\begin{equation}
\Delta(t)= \frac{E(t)^2}{4 U_{po}}- \omega(t)^2. 
 \label{eq:eq4}
\end{equation}
According to the equation (\ref{eq:eq4}), if $\Delta (t)$ is negative then $ U_p(t) < U_{po}$. On the other hand, if $\Delta (t)$ is positive then $U_{po} > U_p(t)$. Consequently, it could be expected, a priori, that if $\Delta (t)$ is negative, the chirp induced in the pulse will be unable to extend the cut-off. On the contrary, if $\Delta(t)$ is positive, in principle, a cut-off extension could be observed as we will see below. 

In order to complement the above described classical analysis we calculate the harmonic spectra by numerical integration of the TDSE-3D in the length gauge within the dipole approximation. As it is well known the harmonic yield of an atom is proportional to the Fourier transform of the dipole acceleration of its active electron and it can be calculated from the  time propagated electronic wave function. We have used our code which is based on an expansion in spherical harmonics, $Y_{l}^{m}$, considering only the $m = 0$ terms due to the cylindrical symmetry of the problem. The numerical technique to solve the TDSE-3D is based on a Crank-Nicolson method implemented on a splitting of the time-evolution operator that preserves the norm of the wave function. Here we base our studies in the helium atom due to the fact that a majority of experiments in HHG are carried out in noble gases. Hence we have considered in our TDSE-3D code the atomic potential reported in~\cite{lin05} to accurately describe the level structure of the helium atom under the Single Active Electron (SAE) approximation.
In addition, and in order to explore the detailed spectral and temporal behaviour of HHG, we perform a time-frequency analysis of the HHG spectra by means of a wavelet transform~\cite{waveletchui,waveletantoine,waveletong}.

\section{Results and discussion}

\subsection{Linear Dispersive Chirp}

The first study of HHG within this context was performed with a chirped pulse induced by a dispersive medium. As it is well known, when an optical pulse passes through a dispersive medium it suffers a temporal broadening~\cite{18}. According to the energy conservation, the area under the pulse must remain constant. Consequently, the peak pulse amplitude has to decrease in order to conserve the laser pulse energy. In a dispersive medium the spectral content of the travelling pulse is not modified during the pulse propagation through the medium. The pulse is, however, temporally stretched. This effect can be explained in a first approximation by expanding the temporal phase in power series and considering the dispersive effect through the GVD term~\cite{19}. This term is defined as, $\mathrm{GVD}=\frac{d^2k}{d\omega^2}$. Introducing the parameter $a = \frac{1}{2}\frac{d^2k}{d\omega^2}$, the evolution of the temporal broadening of a Gaussian pulse propagating a distance $L$ through the medium can be expressed as:
\begin{equation}
\tau(L)= \tau_0\sqrt{1+\left(\frac{8aL\ln(2)}{\tau_0^2}\right)^2},
 \label{eq:eq5}
\end{equation}
wßþhere $\tau_0$ is the initial FWHM. Consequently, the degree of the chirped pulse can be expressed as a function of the product $aL$, whose unit is fs$^2$~\cite{mexicagvd}.    
\begin{figure}[h]
\centering
\includegraphics[width=\textwidth]{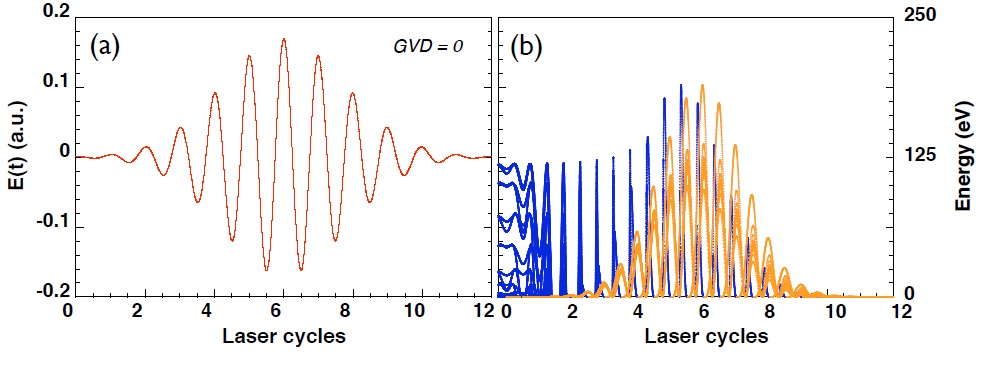}
\caption{Panel (a) Driving laser pulse without chirp, i.e.~GVD=0, at $1\times10^{15}$ W/cm$^2$ of laser peak intensity (i.e. $E_0=0.1688$ a.u.) and $\lambda=800$ nm. Panel (b) represents the corresponding classical energies, resulting from the integration of the Newton-Lorentz equation, at the recombination time as a function of the ionization time (in blue) and recombination time (in orange) (for more details see e.g~\cite{classtrong}). The maximum photon energy, following (\ref{eq:eq1}), is 189 eV (note we have put the origin in energy as the $I_p$ value).}
\label{fig:fig1} 
\end{figure}
For a linear dispersive medium it is feasible to use an ordinary glass type where the GVD parameter is calculated for $\lambda=800$ nm~\cite{20}. Note that at this wavelength one laser period corresponds to $\approx2.6$ fs. The temporary broadening is then calculated for different beams propagating through the dispersive medium for several pathways lengths. We will start our analysis analysing the classical electron energy limits. Figure 1(a) shows a driving laser pulse without chirp, i.e.~GVD=0, at a laser peak intensity of $1\times10^{15}$ W/cm$^2$ (i.e.~$E_0=0.1688$ a.u.) and $\lambda=800$ nm joint with the corresponding classical trajectory analysis, figure 1(b), extracted by the integration of the Newton-Lorentz equation, neglecting the effect of the magnetic field (for more details about the classical simulations see e.g~\cite{classtrong}). In figure~\ref{fig:fig2} three chirped laser pulses for different values of $aL$ are plotted, joint with the classical electron energy simulations. Figure~\ref{fig:fig2}(a) is for $aL=26.6$ fs$^2$, figure~\ref{fig:fig2}(b) for $aL=53.3$ fs$^2$ and figure~\ref{fig:fig2}(c) for $aL=88.8$ fs$^2$, respectively. Furthemore, figures~\ref{fig:fig2}(d), \ref{fig:fig2}(e) and \ref{fig:fig2}(f) represent the corresponding $\Delta(t)$ functions defined by equation (\ref{eq:eq4}). Note that in all cases the $\Delta(t)$ never takes positive values, consequently these linear chirped pulses are not able to increase the harmonic cut-off (see figure 1). This fact is confirmed by the classical analysis as it is shown in figures~\ref{fig:fig2}(g)-\ref{fig:fig2}(i).
\newpage
\begin{figure}[h]
\centering
\includegraphics[width=\textwidth]{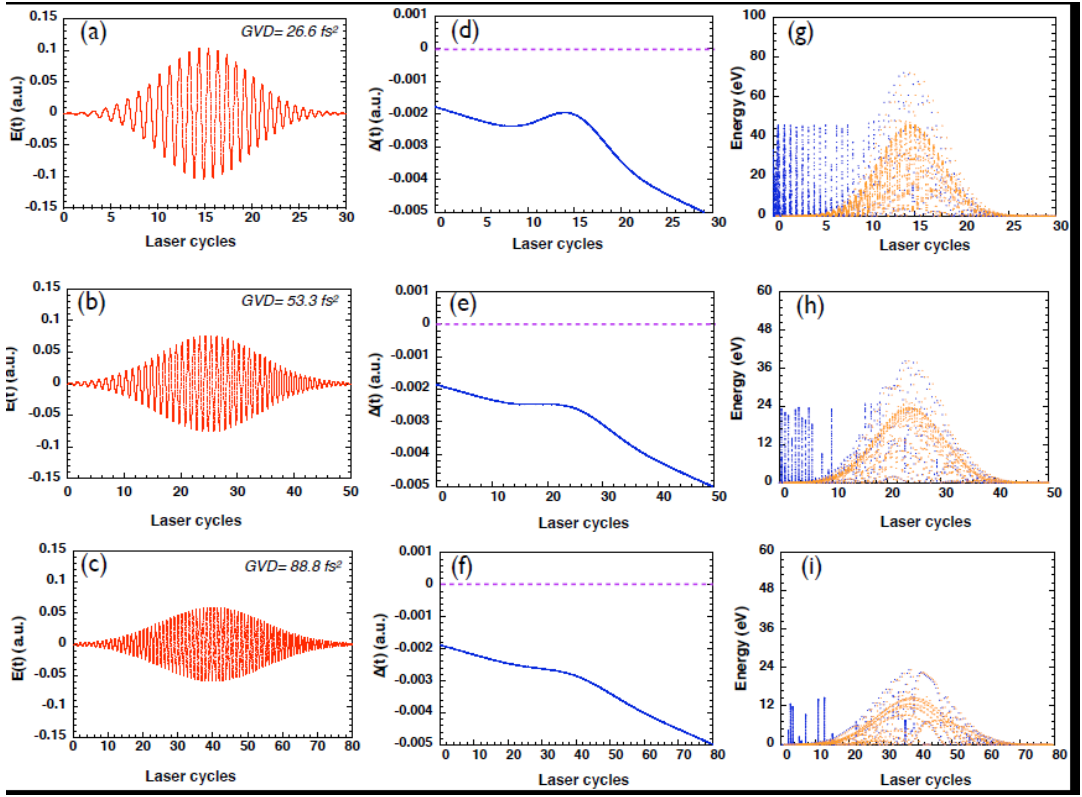}
\caption{Panels (a), (b) and (c) represent the driven chirped laser pulses for the same laser parameters (intensity and wavelength) that in figure~\ref{fig:fig1} but in this case for three different values of the quantity $aL$. The corresponding values of the $\Delta(t)$ function are plotted in (d), (e) and (f), respectively. Following the same criterion of figure~\ref{fig:fig1} panels (g), (h) and (i) represent their respective classical analysis.}
\label{fig:fig2} 
\end{figure} 
\vspace{-.75cm}
\subsection{Non linear chirp}
Our second analysis was performed by assuming that the driving laser pulse is given by the following analytical form,
\begin{equation}
 E(t)=E_0\exp{\left[-2\ln(2)\left(\frac{t}{\tau_0}\right)^2\right]} \cos(\omega_0t+b t^2)
 \label{eq:eq6}
\end{equation}
where $E_0$ is the laser electric field peak amplitude, $\tau_0$ is the FWHM, and $\omega_0$ is the central frequency. The parameter $b$ in equation (\ref{eq:eq6}) determines the degree of chirping. In this case we assume that the pulse envelope does not change, consequently, the maximum field amplitude and the temporal width remain invariant in any case. In the following we will study three particular cases of chirped pulses varying the $b$ parameter. 

\begin{figure}[h]
\centering
\includegraphics[width=\textwidth]{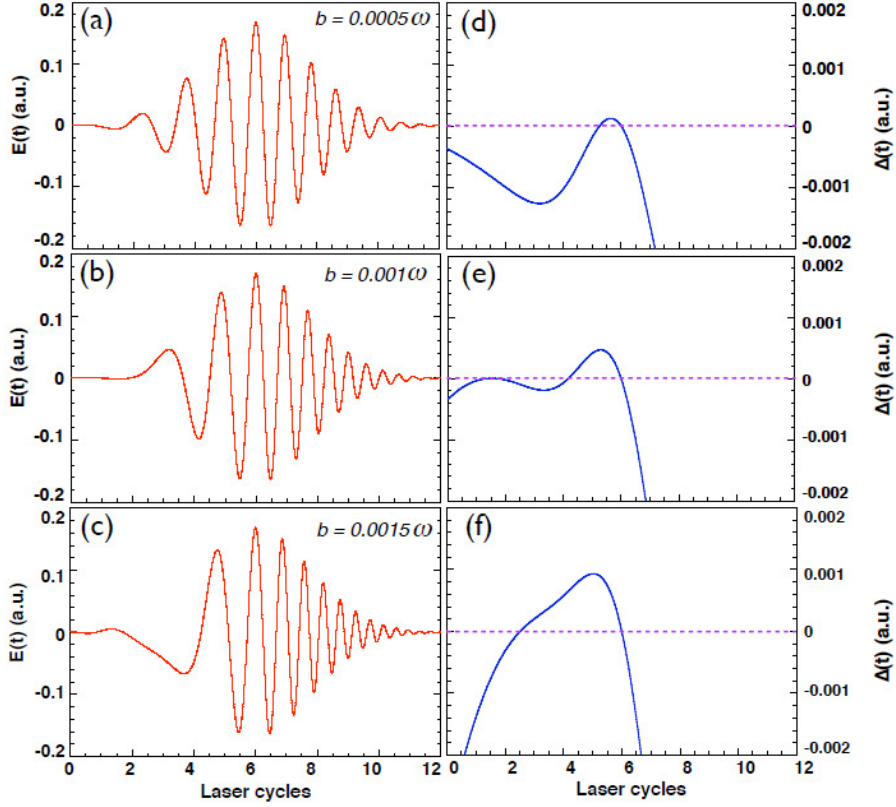}
\caption{Driving chirped laser pulses described by the equation (\ref{eq:eq6}) at the intensity of $1\times10^{15}$ W/cm$^2$ for different values of the parameter $b$. Panel (a) $b=0.0005\omega_0$, panel (b) $b=0.001\omega_0$ and panel (c) $b=0.0015\omega_0$ respectively. Panels (d), (e) and (f) represent the corresponding $\Delta(t)$ function for each case.}
\label{fig:fig3} 
\end{figure}

In figure~\ref{fig:fig3} we plot three driving laser pulses, described by the equation (\ref{eq:eq6}), for different values of $b$. Figure \ref{fig:fig3}(a) is for $b=0.0005\omega_0$, figure \ref{fig:fig3}(b) for $b=0.001\omega_0$ and figure \ref{fig:fig3}(c) for $b=0.0015\omega_0$, respectively. The corresponding $\Delta(t)$ function is plotted in figures \ref{fig:fig3}(d)-(f), respectively. Note that in this case $\Delta(t)$ takes positive values for certain temporal regions along the pulse. 

\begin{figure}[h]
\centering
\includegraphics[width=.6\textwidth]{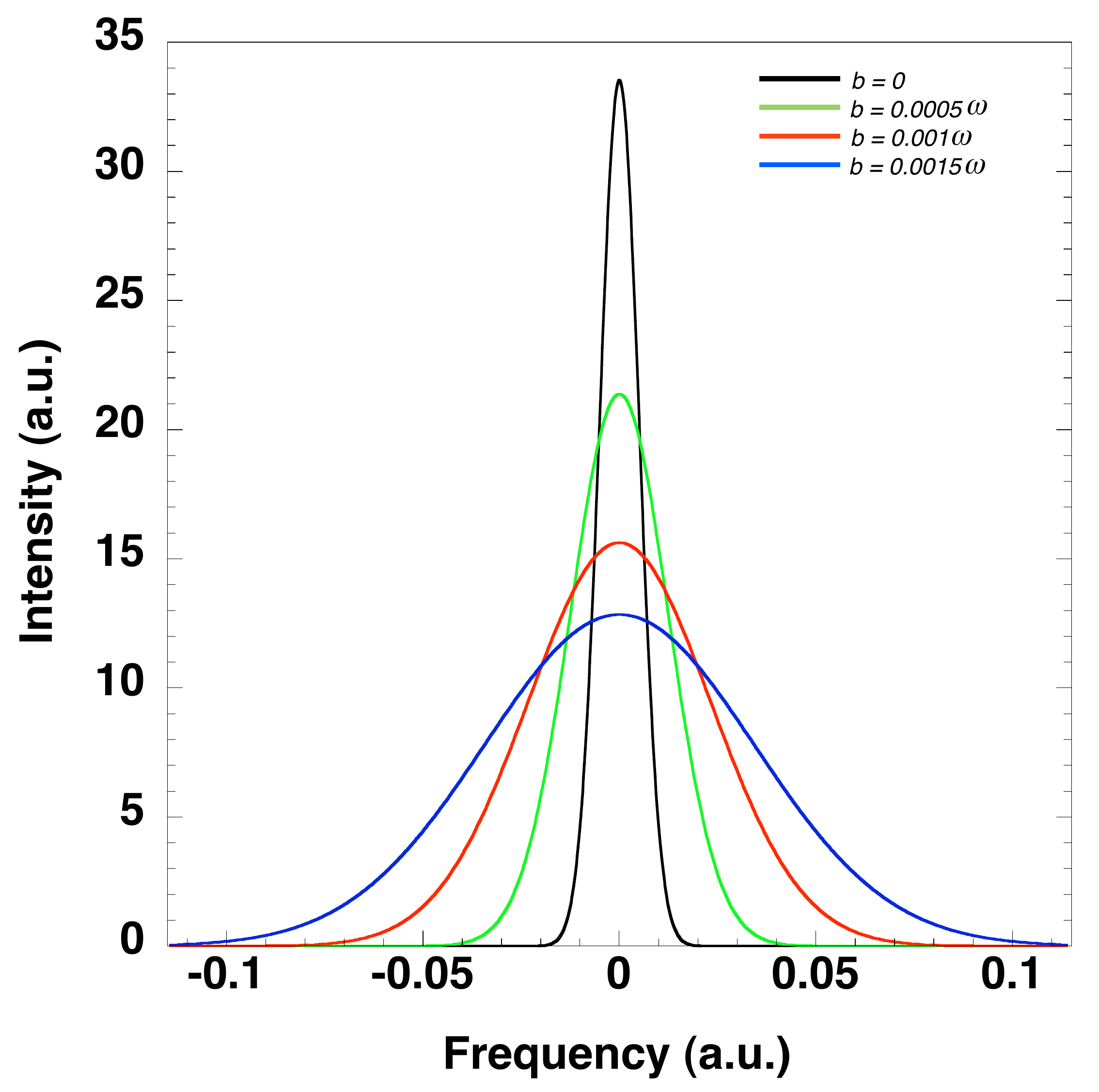}
\caption{Fourier Transform of the corresponding driving laser pulses plotted in figure~\ref{fig:fig3} ($b=0.0005\omega_0$ in green, $b=0.001\omega_0$ in red and $b=0.0015\omega_0$ in blue). The Fourier Transform of the driving pulse without chirp ($b=0$) is plotted in black.}
\label{fig:fig4}
\end{figure}

Figure \ref{fig:fig4} shows one of the key points of this work. In this figure we show that if the $b$ parameter  increases, the pulse spectrum broaden. This spectral broadening allow us to obtain new frequencies, lower and higher than $\omega_0$. As a consequence it will be possible to manipulate the harmonic cut-off in agreement with the cut-off law reported in equation (\ref{eq:eq1}).  

\newpage

\begin{figure}[h]
\centering
\includegraphics[width=10cm]{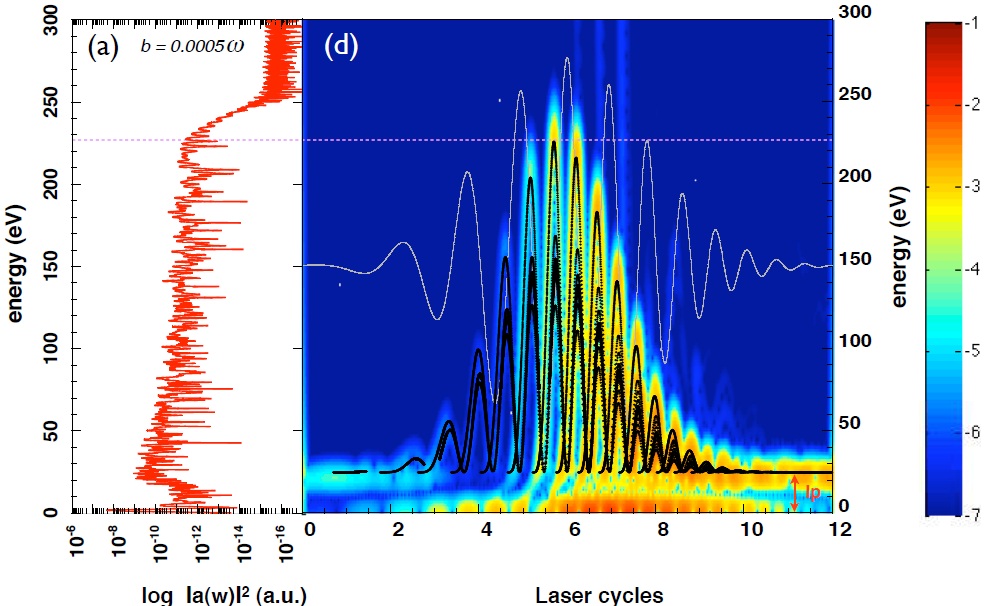}
\includegraphics[width=10cm]{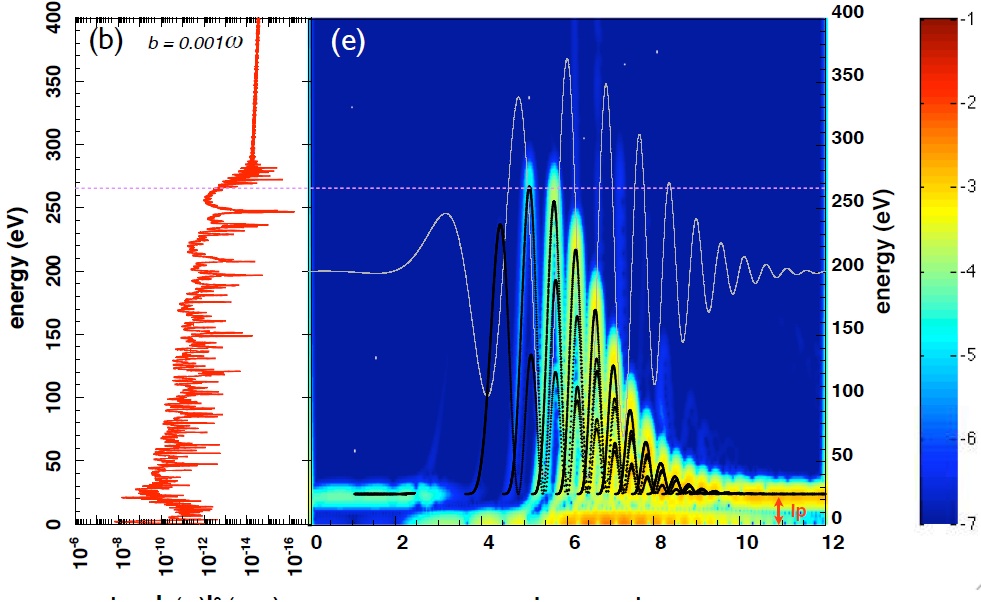}
\includegraphics[width=10cm]{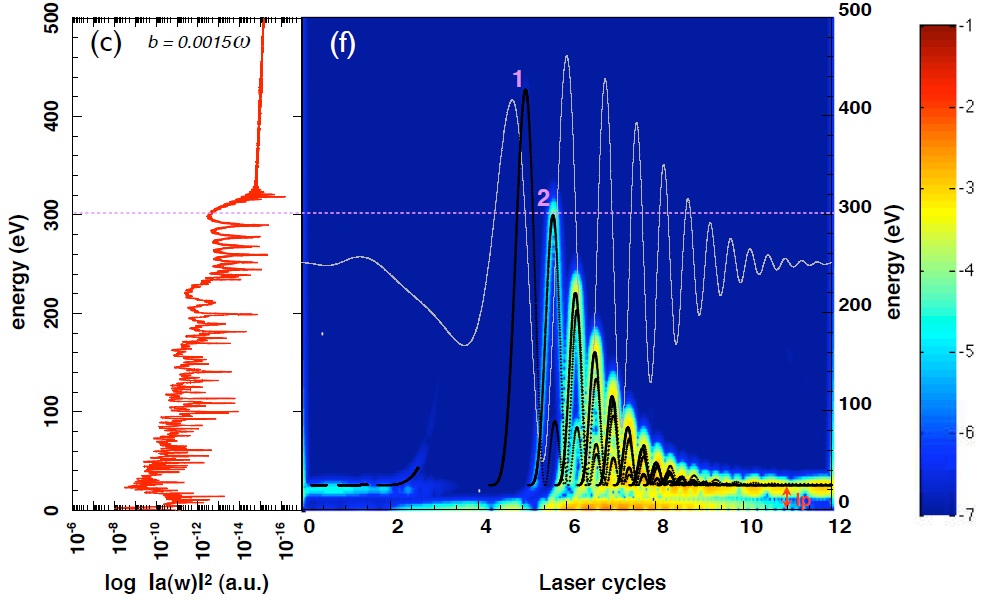}
\caption{Panels (a), (b) and (c) represent the corresponding HHG spectra in the helium atom for the three different chirped laser pulses plotted in the figure~\ref{fig:fig3}. In (d), (e) and (f) are plotted the respective time-frequency analysis, and superimposed (in solid black lines) the classical recombination energies. The laser field is plotted in solid gray. Note that the ionization potential ($I_p$) of the target atom, in this case $I_p=24.7$ eV, has been now included.}
\label{fig:fig5} 
\end{figure}

In the next we will use the TDSE-3D in order to compute the harmonic spectrum in a helium atom. In addition we will compare these quantum mechanical predictions with classical simulations. In figure \ref{fig:fig5}(a), (b) and (c) are plotted the HHG spectra computed with the TDSE-3D. The driven laser pulses are the ones of the figure \ref{fig:fig3}. The time-frequency analysis is shown in figures \ref{fig:fig5}(d), (e) and (f), respectively. From these figures it is possible to account the instant when the harmonics are emitted along the laser pulse. In addition, the classical recombination energies (in solid black lines) have been superimposed. By simple inspection of this figure it is easy to conclude that the quantum simulations fully confirm the cut-off extensions predicted by the classical analysis. 

\newpage

As figure~\ref{fig:fig5} shows, all the cases present an increment of the harmonic cut-off in the single atom response. Both classical and quantum analysis show an excellent degree of accuracy and are in a complete agreement with the predictions reported by the behavior of $\Delta(t)$ (figure 3). The classical analysis confirms that the maximum of recombination energy coincides with the interval in which $\Delta(t)$ is positive. Note that, however, in the case of  $b=0.0015\omega_0$ plotted in figure~\ref{fig:fig5}(c), in spite of the fact that an important cut-off enhancement is achieved, the maximum recombination energy reported by the classical analysis, labelled by the {\em point $1$} does not generate harmonics as figure~\ref{fig:fig5}(c) shows. This is so because the peak amplitude of the corresponding maximum of the field responsible of this recombination event, which correspond to the first maximum at the beginning of the laser pulse in the turn-on region, is so weak, around $0.06$ a.u., i.e.~$1.2\times10^{14}$ W/cm$^2$, to produce tunnel ionization in helium, and consequently unable to generate harmonic radiation. To overcome this limitation we increase the peak laser intensity up to $1.4\times10^{15}$ W/cm$^2$ for the case of $b=0.0015\omega_0$ . 

\begin{figure}[h]
\centering
\includegraphics[width=\textwidth]{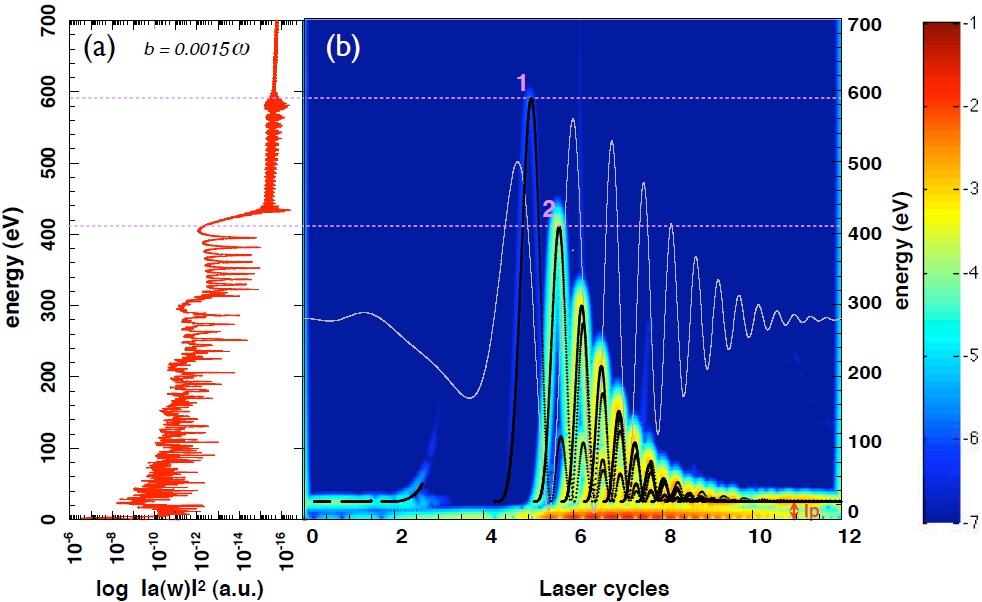}
\caption{Idem figure 5(c) but in this case the peak laser intensity is $1.4\times10^{15}$ W/cm$^2$ ($E_0=0.2$ a.u.) corresponding to the threshold of saturation of helium atom.}
\label{fig:fig6}
\end{figure}

We plot these results in figure~\ref{fig:fig6}. On this way, it is expecting that now the peak field of the first maximum mentioned above, responsible of the maximum electron energy at recombination, turns to be intense enough to produce tunnel ionization and consequently, harmonic radiation. This fact is depicted by the figure~\ref{fig:fig6}(a) where an important cut-off extension appears in agreement by the classical predictions. Note that the new different plateaus generated can be understood considering that the harmonic efficiency depend on $\lambda^{-5.5}$~\cite{22,joseoptexp}. In other words, the generation of each new plateau can be interpreted as a result of ionization events with different wavelengths and for this reason the efficiency of the plateau in the region between 400-600 eV is so poor.

\section{Conclusions and outlook}

We have studied the behavior of the harmonic cut-off in the HHG process driving by two different kind of chirped laser pulses. In the first case we employ a chirped pulse whose spectral content has remained unaltered when it passes through a dispersive media~\cite{21}. In the second case, we have considered a chirped pulse whose frequency varies linearly in time and its envelope remains constant. In such a pulse, and by analysing the harmonic spectrum, it was demonstrated that, as the control chirp parameter $b$ increases, the laser spectrum broaden. 

The effect of chirped pulses on the harmonic cut-off energy obtained with a dispersive linear media has been evaluated by a classical analysis. As can be observed in figure \ref{fig:fig2} when the chirping of the laser pulse increases, for different values of the GVD, its temporal width grows and the maximum cutoff photon energy diminishes. Considering that the pulse is temporally stretched, the maximum electric field peak amplitude decreases. Consequently the associated drop in the laser intensity causes a reduction of the cutoff energy. On the other hand, the $\Delta(t)$ function previously defined in equation (\ref{eq:eq4}) provides a practical tool to predict, a priori, the behavior of the harmonic cut-off in the case of chirped pulses. 


The next study of the cut-off has been performed using the chirped pulses described by the equation (\ref{eq:eq6}) and plotted in figure \ref{fig:fig3}. As can be observed, these pulses have a chirp whose frequency depends linearly in time but its temporal duration remain constant in spite of variations of $b$, the parameter that controls the chirp. Analysing the behavior of $\Delta(t)$ it is possible to correlate the increasing in the harmonic cut-off with regions in which $\Delta(t)>0$. This conclusion is in perfect agreement with the simple analysis based on the ponderomotive energy presented in Section 2.

 
In summary, according to the results reported here, chirping laser pulses will be able to produce cut-off enhancements only in the case that the chirp is obtained via a non linear process where new frequencies are generated and a broadening of the laser spectrum is achieved~\cite{21}. In particular, lower frequencies, i.e.~longer wavelengths, than the central frequency are the ones which produce the most important cutoff extensions. This behavior can be understood based on the simple law $E_{cut-off}\propto I\lambda^2$. These new frequencies are generated by weak laser peak amplitudes, at the turn-on region of the driving pulse. Consequently, as an additional condition needed in order to produce noticeable effects in the HHG spectra, the input pulse have to be intense enough to produce tunnel ionization, and consequently harmonic radiation, at the turn-on temporal region.

\section*{Acknowledgements}
This work was partially supported by Agencia de Promoci\'{o}n Cient\'{i}fica y Tecnol\'{o}gica (Argentina) under project PICT-2010-2575. J.A.P.-H. and L.R. acknowledge support from Laserlab-Europe (Grant No.~EU FP7 284464) and the Spanish Ministerio de Econom\'ia y Competitividad (FURIAM Project No. FIS2013-47741-R).


\section*{References}

\begin{thebibliography}{30}

\bibitem{1} L'Huillier A and Balcou Ph 1993 High-order harmonic generation in rare gases with a 1-ps 1053-nm laser \textit{Phys. Rev. Lett.} \textbf{70} 774 

\bibitem{2} Macklin J J, Kmetec J D and Gordon C L 1993 High-order harmonic generation using intense femtosecond pulses \textit{Phys. Rev. Lett.} \textbf{70} 766

\bibitem{3} Corkum P B 1993 Plasma perspective on strong field multiphoton ionization \textit{Phys. Rev. Lett.} \textbf{71} 1994

\bibitem{4} Krause J L, Schafer K J and Kulander K C 1992 High-order harmonic generation from atoms and ions in the high intensity regime \textit{Phys. Rev. Lett.} \textbf{68} 3535

\bibitem{5} Schafer K J, Yang B, DiMauro L F and Kulander K C 1993 Above threshold ionization beyond the high harmonic cutoff \textit{Phys. Rev. Lett.} \textbf{70} 1599 

\bibitem{maciej} Lewenstein M, Balcou Ph, Ivanov M Yu, L'Huillier A and Corkum P B 1994 Theory of high-harmonic generation by low-frequency laser fields \textit{Phys. Rev. A} \textbf{49} 2117

\bibitem{amelleprx} Haessler S, Bal\u{c}iunas T, Fan G, Andriukaitis G, Pug\u{z}lys A, Baltu\u{s}ka A, Witting T, Squibb R, Za\"{i}r A, Tisch J W G, Marangos J P and Chipperfield J E 2014 Optimization of Quantum Trajectories Driven by Strong-Field Waveforms \textit{Phys. Rev. X} \textbf{4} 021028

\bibitem{vozzi1} Calegari F, Lucchini M, Kim K S, Ferrari F, Vozzi C, Stagira S, Sansone G and Nisoli M 2011 Quantum path control in harmonic generation by temporal shaping of few-optical-cycle pulses in ionizing media \textit{Phys. Rev. A} \textbf{84} 041802(R)

\bibitem{6} Winterfeldt C, Spielmann C and Gerber G 2008 Colloquium: Optimal control of high-harmonic generation \textit{Rev. Mod. Phys.} \textbf{80} 117 

\bibitem{7} Pfeifer T, Kemmer R, Spitzenpfeil R, Walter D, Winterfeldt C, Gerber G and Spielmann C 2005 Spatial control of high-harmonic generation in hollow fibers \textit{Opt. Letters} \textbf{30} 1497

\bibitem{optexp2012} Ciappina M F, A\'cimovi\'c S S, Shaaran T, Biegert J, Quidant R and Lewenstein M 2012 Enhancement of high harmonic generation by confining electron motion in plasmonic nanostrutures \textit{Opt. Exp.} \textbf{20} 26261 

\bibitem{joseprl} P\'erez-Hern\'andez J A, Ciappina M F, Lewenstein M, Roso L and Za\"ir A 2013 Beyond Carbon K-edge harmonic emission using spatial and temporal synthesized laser fields \textit{Phys. Rev. Lett.} \textbf{110} 053001

\bibitem{classtrong} Ciappina M F, P\'erez-Hern\'andez J A and Lewenstein M 2014 CLAssSTRONG: Classical simulations of strong field processes \textit{Comp. Phys. Comm.} \textbf{185} 398

\bibitem{atispra} Ciappina M F, P\'erez-Hern\'andez J A, Shaaran T, Roso L and Lewenstein M 2013 Electron-momentum distributions and photoelectron spectra of atoms driven by an intense spatially inhomogeneous field \textit{Phys. Rev. A} \textbf{87} 063833

\bibitem{atislpl} Ciappina M F, Shaaran T, Guichard R, P\'erez-Hern\'andez J A, Roso L, Arnold M, Siegel T, Za\"ir A and Lewenstein M 2013 High energy photoelectron emission from gases using plasmonic enhanced near-fields \textit{Laser Phys. Lett.} \textbf{10} 105302 

\bibitem{8} Kim H T, Kim I J, Tosa V, Kim C M, Park J J, Lee Y S, Bartnik A, Fiedorowicz H and Nam C H 2004 Bright High-Order Harmonic Generation From Long Gas Jets Toward Coherent Soft X-Ray Applications \textit{IEEE J. Sel. Top. Quantum Electron.} \textbf{10} 1329

\bibitem{9} Kim H T, Kim I J, Lee D G, Hong K-H, Lee Y S, Tosa V and Nam C H 2004 Optimization of high-order harmonic brightness in the space and time domains \textit{Phys. Rev. A} \textbf{69} 031805

\bibitem{10} Lee D G, Kim J-H, Hong K-H and Nam C H 2001 Coherent Control of High-Order Harmonics with Chirped Femtosecond Laser Pulses \textit{Phys. Rev. Lett.} \textbf{87} 243902

\bibitem{11} Kim H T, Kim I J, Hong K-H, Lee D G, Kim J-H and Nam C H 2004 Chirp analysis of high-order harmonics from atoms driven by intense femtosecond laser pulses \textit{J. Phys. B} \textbf{37} 1141
 
\bibitem{12} Kim H T, Lee D G, Hong K-H, Kim J-H, Choi I W and Nam C H 2003 Continuously tunable high-order harmonics from atoms in an intense femtosecond laser field \textit{Phys. Rev. A} \textbf{67} 051801

\bibitem{13} Ultrashort Laser Pulse Phenomena: Fundamentals, Techniques, and Applications on a Femtosecond Time Scale, 2nd ed., by Jean-Claude Diels and Wolfgang Rudolph Ultrashort Laser Pulse Phenomena: Fundamentals, Techniques, and Applications on a Femtosecond Time Scale, 2nd ed. Jean-Claude Diels and Wolfgang Rudolph , Elsevier, New York, 2006 , 82 pp. 

\bibitem{14} Carrera J J and Chu S-I 2007 Extension of high-order harmonic generation cutoff via coherent control of intense few-cycle chirped laser pulses \textit{Phys. Rev. A} \textbf{75} 033807

\bibitem{15} Wu J, Zhang G-T, Xia C-L and Liu X-S 2010 Control of the high-order harmonics cutoff and attosecond pulse generation through the combination of a chirped fundamental laser and a subharmonic laser field \textit{Phys. Rev. A} \textbf{82} 013411

\bibitem{16} Du H and Hu B 2011 Propagation effects of isolated attosecond pulse generation with a multicycle chirped and chirped-free two-color field \textit{Phys. Rev. A} \textbf{84} 023817 

\bibitem{17} Niu Y, Xiang Y, Qi Y and Gong S 2009 Single attosecond pulse generation from multicycle nonlinear chirped pulses \textit{Phys. Rev. A} \textbf{80} 063818

\bibitem{lin05} Tong X M and Lin C D 2005 Empirical formula for static field ionization rates of atoms and molecules by lasers in the barrier-suppression regime \textit{J. Phys. B: At. Mol. Opt. Phys.} \textbf{38} 2593

\bibitem{waveletchui} Chui C K 1992 An Introduction to Wavelets Academic Press New York

\bibitem{waveletantoine} Antoine A Piraux B and Maquet A 1995 Time profile of harmonics generated by a single atom in a strong electromagnetic field \textit{Phys. Rev. A} \textbf{51}, 1750(R)

\bibitem{waveletong} Tong X M and Chu S-I 2000  Probing the spectral and temporal structures of high-order harmonic generation in intense laser pulses \textit{Phys. Rev. A}  \textbf{61}, 021802(R) 
			                
\bibitem{gaard01A} Gaarde M B Time-frequency representations of high order harmonics 2001 \textit{Opt. Express} \textbf{8}, 529

\bibitem{18} Ultrashort Laser Pulse Phenomena: Fundamentals, Techniques, and Applications on a Femtosecond Time Scale, 2nd ed., by Jean-Claude Diels and Wolfgang Rudolph Ultrashort Laser Pulse Phenomena: Fundamentals, Techniques, and Applications on a Femtosecond Time Scale, 2nd ed. Jean-Claude Diels and Wolfgang Rudolph , Elsevier, New York, 2006 , 19 pp. 

\bibitem{19} Ultrashort Laser Pulse Phenomena: Fundamentals, Techniques, and Applications on a Femtosecond Time Scale, 2nd ed., by Jean-Claude Diels and Wolfgang Rudolph Ultrashort Laser Pulse Phenomena: Fundamentals, Techniques, and Applications on a Femtosecond Time Scale, 2nd ed. Jean-Claude Diels and Wolfgang Rudolph , Elsevier, New York, 2006 , 16 pp. 

\bibitem{mexicagvd} M Rosete-Aguilar  Calculation of temporal spreading of ultrashort pulse propagating through optical glasses 2008 \textit{Rev. Mex. Fis.}  \textbf{54}, 141-148


\bibitem{20} Ultrashort Laser Pulse Phenomena: Fundamentals, Techniques, and Applications on a Femtosecond Time Scale, 2nd ed., by Jean-Claude Diels and Wolfgang Rudolph Ultrashort Laser Pulse Phenomena: Fundamentals, Techniques, and Applications on a Femtosecond Time Scale, 2nd ed. Jean-Claude Diels and Wolfgang Rudolph , Elsevier, New York, 2006 , 44 pp. 

\bibitem{21} Ultrashort Laser Pulse Phenomena: Fundamentals, Techniques, and Applications on a Femtosecond Time Scale, 2nd ed., by Jean-Claude Diels and Wolfgang Rudolph Ultrashort Laser Pulse Phenomena: Fundamentals, Techniques, and Applications on a Femtosecond Time Scale, 2nd ed. Jean-Claude Diels and Wolfgang Rudolph , Elsevier, New York, 2006 , 29 pp. 

\bibitem{22} Tate J, Auguste T, Muller H G, Sali\`eres P, Agostini P, and DiMauro L F 2007 Scaling of wave-packet dynamics in an intense midinfrared field \textit{Phys. Rev. Lett.} \textbf{98} 013901

\bibitem{joseoptexp} P\'erez-Hern\'andez J A, Roso L, and Plaja L 2009 Harmonic generation beyond the Strong-Field Approximation: the physics behind the short-wave-infrared scaling laws \textit{Opt. Exp.} \textbf{17} 9891
  


\end{thebibliography}
\end{document}